\documentclass[letterpaper, 10 pt, conference]{ieeeconf}
\IEEEoverridecommandlockouts
\usepackage{graphicx} 
\usepackage{float}

\usepackage{epstopdf} 
\usepackage{epsfig}
\usepackage{mathptmx} 
\usepackage{times} 
\usepackage{amsmath} 
\usepackage{amssymb}  
\usepackage{mathtools}
\usepackage{cite}
\usepackage[final]{pdfpages}
\usepackage{color} 
\usepackage{siunitx} 
\usepackage{dblfloatfix}
\let\DeclareUSUnit\DeclareSIUnit

\DeclareUSUnit\inch{"}

\usepackage{textcomp}

\usepackage{multirow}
\usepackage{array}
\newcolumntype{P}[1]{>{\centering\arraybackslash}p{#1}}

\usepackage[utf8]{inputenc}
\usepackage{comment} 
\usepackage{array}

\usepackage{hyperref}

\usepackage{array}
\newcolumntype{C}[1]{>{\centering\arraybackslash}p{#1}}

\title{\LARGE \bf A Computational Design and Evaluation Tool\\for 3D Structures with Planar Surfaces}

\author{Chang Liu$^{\dag}$, Wenzhong Yan, Pehu\'en Moure, Cody Fan, and Ankur Mehta
\thanks{Chang Liu, Pehu\'en Moure, Cody Fan and Ankur Mehta are with the Samueli School of Engineering, Electrical and Computer Engineering, Wenzhong Yan is with the Samueli School of Engineering, Mechanical and Aerospace Engineering, University of California, Los Angeles, CA, USA \newline
\indent $^{\dag}$ Corresponding author, {\tt\footnotesize changliu498@ucla.edu}
\newline
\indent This work is supported by the National Science Foundation (\#1752575).
\newline
\indent Programming package can be found at \href{https://git.uclalemur.com/Chang_Liu/Identifying_Design_Weakness_In_Foldabe_Structures/-/tree/master/2_Modeling_Coding/_Design_Tool_Final}{\textit{\textcolor{black}{https://git.uclalemur.com/
Chang\_Liu/Identifying\_Design\_Weakness\_In\_Foldabe\_Structures/-/tree/master/2\_Modeling\_Coding/\_Design\_Tool\_Final}}}
}
}

\begin{document}

\maketitle



\begin{abstract}


Three dimensional (3D) structures composed of planar surfaces can be build out of accessible materials using easier fabrication technique with shorter fabrication time. To better design 3D structures with planar surfaces, realistic models are required to understand and evaluate mechanical behaviors. Existing design tools are either effort-consuming (e.g. finite element analysis) or bounded by assumptions (e.g. numerical solutions). In this project, We have built a computational design tool that is (1) capable of rapidly and inexpensively evaluating planar surfaces in 3D structures, with sufficient computational efficiency and accuracy; (2) applicable to complex boundary conditions and loading conditions, both isotropic materials and orthotropic materials; and (3) suitable for rapid accommodation when design parameters need to be adjusted. We demonstrate the efficiency and necessity of this design tool by evaluating a glass table as well as a wood bookcase, and iteratively designing an origami gripper to satisfy performance requirements. This design tool gives non-expert users as well as engineers a simple and effective modus operandi in structural design.
\end{abstract}



\section{Introduction}
\label{section:intro}

3D structures built by combining planar parts are extensively applied in various aspects, such as in our daily life \cite{bezzo2015robot,umetani2012guided}, architectures \cite{gilewski2014origami}, nature\cite{Luque2002}, and engineering\cite{rus2018design,liu2017self,liu2019self}. For example, engineers build light-weight 3D structures and robots from 2D planar materials by harnessing origami-inspired folding as a design and fabrication method \cite{liu2019self,yan2018towards}.
This technique has gained its popularity due to its advantages: (i) accessible materials, e.g. plywood, cardboard, and paper; (ii) ease of fabrication, e.g. using saws and paper cutter; (iii) shorter building time and less required training compare to other 3D manufacturing techniques, e.g. 3D printing, milling, lathing and drilling; and (iv) suitable for building large scale and small scale structures.

\begin{figure}[t]
	\includegraphics[width=3.25in]{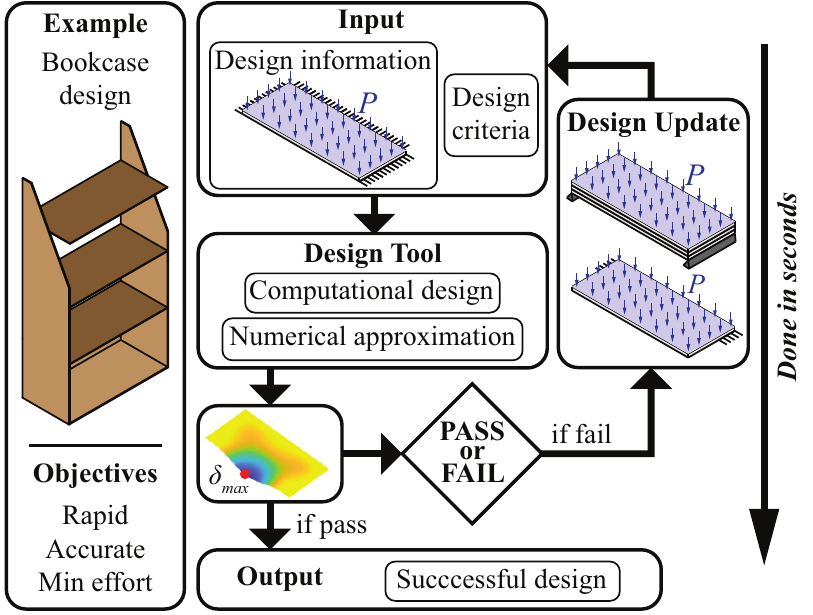}
	\centering
	\caption{A flow chart of our computational design paradigm with a bookcase design example.}
	\label{fig:Introduction}
\end{figure}

Even though 3D structures with planar surfaces have such potentials as mentioned above, designing them is still quite challenging.
The design of 3D structures with planar surfaces requires iterative approach for accruing satisfactory performance, which can be both consumptive and laborious.
This necessitates a demand for design tools that can both efficiently aid non-expert users to evaluate designs based on desired behaviors as well as facilitate engineers to perform effective iterative design at preliminary design stages.

However, exact solutions for analyzing 3D structures can be complex and sometimes impossible to obtain. For simplification analysis, previous research have analyzed planar surfaces in 3D structures individually and then incorporated them through hinges (e.g. origami devices \cite{filipov2017bar} and paneled furniture \cite{eckelman1967furniture}). This provides us an insight to investigate planar surfaces in 3D structures individually as our evaluation approach in our design tool.

There exist some evaluation tools that have been used on planar surfaces in 3D structures:
(1) There  exist various analytical solutions in solving plate problems \cite{timoshenko2009theory}. Even though analytical solutions give the exact results, existing solutions are few and can only analyze limited plate shapes, boundary conditions and loading conditions.
(2) Finite Element Analysis (FEA) has been widely used in various engineering fields. However, FEA can be cumbersome, with uncertain accuracy and speed \cite{woodruff2020bar}. It is time consuming and computationally expensive, therefore not a suitable tool at preliminary design stages, especially for non-expert users.
(3) There are other numerical solutions for approximating plate problem, such as the Finite Difference Method (FDM), the Boundary Collocation Method (BCM), the Boundary Element Method (BEM), the Galerkin Method, and the Ritz Method \cite{ventsel2002thin}. While numerical solutions are straightforward and simple to use, there are extensive assumptions attached to them, which limits their feasibility \cite{zozulya2011numerical}.
(4) Some mathematical models have also been developed for evaluating planar 3D structures, such as the bar-hinge model, which replaces plates with extensional bars and rotational springs \cite{filipov2015origami}. Even though it is computationally cheap and able to capture the deformations happening in plates, only selected nodes can be calculated accurately \cite{zhu2019simulating}.

Computational design has been proven to be efficient with good accuracy in rapid structural design \cite{liu2020computational, yan2019rapid}.
In this project, we have incorporated existing plate theories with a numerical method (i.e. Galerkin Method of Weighted Residual) to build an effective and straightforward computational design tool that can be easily adopted by non-expert users as well as rapidly used by engineers in the design and evaluation process (Fig.~\ref{fig:Introduction}).

The contributions of this work include the followings:
\begin{itemize} 
    \item a computationally rapid and inexpensive evaluation of 3D structures with planar surfaces;
    \item a tool capable of rapidly predicting structural failure and evaluating potential solutions for iterative design;
    \item demonstrations of our design tool's capabilities and feasibility: enabling structural rapid failure prediction, rapid redesign; easy to adjust design parameters (e.g. geometry information, boundary conditions and loading conditions); and applicable to both isotropic materials and anisotropic materials.
\end{itemize}


The remainder of this paper is organized as follows:
in Section~\ref{section:Background}, we review plate bending theory and Galerkin Method of Weighted Residual;
in Section~\ref{section:Design_Tool_Implementation}, we present the detailed implementation of our design tool using the theory and numerical solution technique reviewed in Section~\ref{section:Background};
the design efficiency and accuracy are demonstrated in Section~\ref{section:Design_Efficiency_Validation};
two furniture examples and an origami gripper example are presented in Section~\ref{section:Design_Implementation};
in the end, the conclusion is discussed in Section~\ref{section:Conclusion}.


\section{Background}
\label{section:Background}

In this section, classical plate bending theory is presented first, followed by method of weighted residual as a numerical method to effectively approximate solutions in this plate bending problem. By applying this approach, one can use our design tool to rapidly and effectively analysis bending planar surfaces in 3D structures.


\subsection{Classical Kirchhoff Plate Bending Theory}

The classical Kirchhoff–Love static plate bending problem assumes: (1) thin plate; (2) small deformation; and (3) constant thickness, etc.
The problem can be expressed as the following governing differential equation \cite{xu2020analytical}:
\begin{equation}
    D_x \frac{\partial^4 \omega}{\partial x^4} + 2 D_{xy} \frac{\partial^4 \omega}{\partial x^2 \partial y^2} + D_y \frac{\partial^4 \omega}{\partial y^4} = \textbf{D}(\omega) = P
\end{equation}
\noindent where $D_x$, $D_y$ and $D_{xy}$ are the bending rigidity of orthotropic plate, $\omega = \omega_{(x, y)}$ is the deflection, \textbf{D} is the differential operator, $P = P_{(x, y)}$ is the distributed transverse loading per unit area. For isotropic material, $D_x = D_y = D_{xy} = D$.


\subsection{Boundary Conditions}

There are three types of boundary conditions \cite{xu2020analytical, he2020bending}:
\begin{itemize}
    \item simply-supported: $\omega=0$ and $M_n=0$;
    \item clamped: $\omega=0$ and $\frac{\partial\omega}{\partial n}=0$;
    \item free: $M_n=0$ and $V_n=0$.
\end{itemize}

\noindent where $M_n$ is the moment normal to the boundary, $n$ is the normal direction to the boundary, and $V_n$ is the reaction force normal to the boundary.


\subsection{Method of Weighted Residuals}

When physical formulation of a problem is described as a differential equation, method of weighted residuals (MWR) is widely used to approximate numerical solutions \cite{finlayson2013method}.
The base function (approximated deflection function $\tilde{\omega}_{(x, y)}$ in this case) and residual function $R_{(x,y)}$ can be expressed as 
\begin{equation}
    \tilde{\omega}_{(x, y)} = \sum_{i=1}^{k} c_i \varphi_{i(x,y)}
    \label{eq:omega}
\end{equation}
\begin{equation}
    R_{(x,y)} = \textbf{D} (\tilde{\omega}_{(x, y)}) - P_{(x,y)}
    \label{eq:residual}
\end{equation}

\noindent where $\varphi_{i(x,y)}$ is shape functions and $c_i$ are coefficients.
The goal is to reduce residual $R_{(x,y)}$ so as to get more accurate results. By applying MWR to plate bending theory, solving fourth order differential equations is reduced to solving linear equations, which is computational inexpensive and rapid.

\section{Design Tool Development}
\label{section:Design_Tool_Implementation}

In this section, we discuss in detail how we develop our design tool by applying MWR to approximate solutions of plate bending problems in a rapid fashion.


\subsection{Loading Condition}
\label{section:Loading_Condition}

In order to ensure easy integration, fast calculation and generalization to various type of loading condition, loads are approximated using polynomial regression \cite{polynomial_regression, Polynomial_Regression_1998}. Fifth order polynomials are used in this project to ensure efficiency while maintaining accuracy.
\begin{equation}
    P_{(x,y)} = \sum_{r+s=0}^{5} b_{rs} x^{r} y^{s}
    \label{eq:loading}
\end{equation}


\subsection{Shape Function Selection}
\label{section: Shape Function Selection}

$n^{th}$ order two-dimensional polynomial transformations are applied as linearly independent sets to build shape function $\varphi_{(x,y)}$ (Eq.~\ref{eq: phi_p}) \cite{2d_poly_transf}. Unknown coefficients $a_j$ are solved by applying $\varphi_{(x,y)}$ to satisfy boundary conditions \cite{xu2020analytical, he2020bending}.
\begin{equation}
    \varphi_{(x,y)} = \sum_{j=1}^{r} a_j x^{p} y^{q}
    ,\hspace{0.3cm}
    p + q \leq n
    ,\hspace{0.3cm}
    r = \frac{(n+1)(n+2)}{2}
    \label{eq: phi_p}
\end{equation}


\subsection{Residual Calculation}
\label{section: Residual Calculation}

Coefficients $c_i$ can be calculated by solving the residual function (Eq.~\ref{eq:residual}).
There are different methods for solving residual functions, such as Galerkin method, least square method, collocation method, and subdomain method \cite{lindgren2009weighted}. Among these methods, Galerkin method is capable of solving problems with more elaborated geometries, reduces the dimensionalilty of the problem faster \cite{narendranath2017board}. Therefore, we choose to use Galerkin method of weighted residual in this project, where residual function is then expressed as
\begin{equation}
    \big( R_{(x,y)}, \varphi_{(x, y)} \big) = 0
    \hspace{0.3cm} \Longrightarrow \hspace{0.3cm}
    \int\int R_{(x,y)} \varphi_{(x, y)} dx dy = 0
    \label{eq: residual_calculation}
\end{equation}


\begin{figure}[t]
	\includegraphics[width=3.25in]{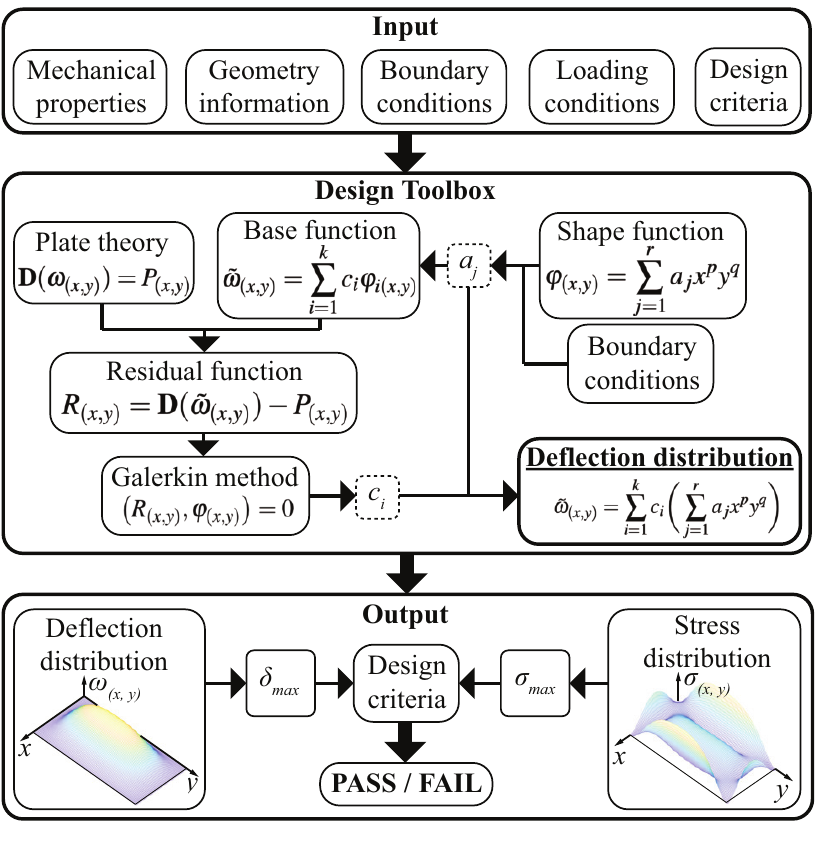}
	\centering
	\caption{Work flow of our computational design tool.}
	\label{fig:Computational_design}
\end{figure}

\subsection{Design Tool Implementation}

\subsubsection{System Input}

There are five inputs in our design tool.

(i) Mechanical properties: Young's modulus $E_{ij}$, Poisson's ratio $\nu_{ij}$, shear modulus $G_{ij}$ and constant thickness $t$. $i$ and $j$ indicate directions in Cartesian system;

(ii) Geometry information: plate vertices locations;

(iii) Boundary conditions: constraint types on edges;

(iv) Loading conditions: uniform loads approximated in polynomial forms as defined in Section~\ref{section:Loading_Condition};

(v) Design criteria: failure modes with metrics, such as industrial standards and/or specified behavioral specifications, etc.


\subsubsection{System Implementation}

As shown in Fig.~\ref{fig:Computational_design}, MWR with Galerkin method is used in our design tool to rapidly approximate deflection solution $\tilde{\omega}_{(x, y)}$ with the minimum polynomial order $n$ automatically determined by our design toolbox.
Once we have all parameters $c_i$'s $a_j$'s resolved by processes stated in Section~\ref{section: Shape Function Selection} and \ref{section: Residual Calculation}, our design tool will generate an approximate polynomial solution for out-of-plane deflection as
\begin{equation}
    \tilde{\omega}_{(x, y)} = \sum_{i=1}^{k} c_i \bigg( \sum_{j=1}^{r} a_j x^{p} y^{q} \bigg)
\end{equation}


\subsubsection{System Output}
Besides the deflection function mentioned above, our design tool is also capable of outputting derivative behavioral properties. Using the approximated polynomial solution, by taking partial differentials, our system can generate other distributions across the plate, such as stresses, moments, shear forces, etc. Those can be further used in evaluating the feasibility of the resulting designs by comparing with specific design criteria.

Thanks to the high speed and low computational consumption, our design tool is desired for rapid iterative design. When the output indicates failure, the system design parameters (e.g. geometry information, boundary conditions, and loading) can be continuously and easily adjusted in a rapid manner until a successful design is gained.


\section{Design Efficiency and Accuracy Validation}
\label{section:Design_Efficiency_Validation}

In this section, we first compare the computational complexity between our design tool and FEA. We then verify our comparison using an example by comparing the speed and accuracy of our design tool (implemented in MATLAB R2019a) with FEA (Solidworks Simulation, 2019 Edition) and an analytical solution.

\begin{table*}[!t]
\caption{Calculation speed and accuracy comparison between Solidworks Simulation and our design tool}
\centering
\begin{tabular}{|C{0.09\textwidth}|C{0.065\textwidth}|C{0.08\textwidth}|C{0.13\textwidth}|C{0.065\textwidth}|C{0.073\textwidth}|C{0.08\textwidth}|C{0.13\textwidth}|C{0.065\textwidth}|}
\hline
\begin{tabular}[c]{@{}c@{}}Analytical\\ solution \cite{imrak2007exact}\end{tabular} & \multicolumn{4}{c|}{Solidworks Simulation} & \multicolumn{4}{c|}{Our design tool} \\ \hline
\begin{tabular}[c]{@{}c@{}}$\omega_{max}$\\($\times10^{-4}\SI{}{\milli\meter}$)\end{tabular} & \begin{tabular}[c]{@{}c@{}}Mesh size\\($\SI{}{\milli\meter}$)\end{tabular} & \begin{tabular}[c]{@{}c@{}}$\omega_{max}$\\($\times10^{-4}\SI{}{\milli\meter}$)\end{tabular} & \begin{tabular}[c]{@{}c@{}}$\Delta \omega_{max}$ ($\times10^{-4}\SI{}{\milli\meter}$)\\ {[}error \%{]}\end{tabular} & \begin{tabular}[c]{@{}c@{}}CPU time\\(sec)\end{tabular} & \begin{tabular}[c]{@{}c@{}}Polynomial\\order n\end{tabular} & \begin{tabular}[c]{@{}c@{}}$\omega_{max}$\\($\times10^{-4}\SI{}{\milli\meter}$)\end{tabular} & \begin{tabular}[c]{@{}c@{}}$\Delta \omega_{max}$ ($\times10^{-4}\SI{}{\milli\meter}$)\\ {[}error \%{]}\end{tabular} & \begin{tabular}[c]{@{}c@{}}CPU time\\(sec)\end{tabular} \\ \hline
\multirow{8}{*}{4.1214} & \multirow{2}{*}{17} & \multirow{2}{*}{0.7510} & \multirow{2}{*}{3.3704 {[}81.78\%{]}} & \multirow{2}{*}{4.78} & \multirow{4}{*}{8} & \multirow{4}{*}{4.0658} & \multirow{4}{*}{0.0556 {[}1.35\%{]}} & \multirow{4}{*}{4.23} \\  &  &  &  &  &  &  &  &  \\ \cline{2-5}
 & \multirow{2}{*}{15} & \multirow{2}{*}{1.1920} & \multirow{2}{*}{2.9294 {[}71.08\%{]}} & \multirow{2}{*}{7.53} &  &  &  &  \\
 &  &  &  &  &  &  &  &  \\ \cline{2-9} 
 & \multirow{2}{*}{6.5} & \multirow{2}{*}{4.0740} & \multirow{2}{*}{0.0474 {[}1.15\%{]}} & \multirow{2}{*}{32.33} & \multirow{4}{*}{12} & \multirow{4}{*}{4.1199} & \multirow{4}{*}{0.0015 {[}0.04\%{]}} & \multirow{4}{*}{7.88} \\
 &  &  &  &  &  &  &  &  \\ \cline{2-5}
 & \multirow{2}{*}{2} & \multirow{2}{*}{4.1230} & \multirow{2}{*}{0.0016 {[}0.04\%{]}} & \multirow{2}{*}{732.14} &  &  &  &  \\ 
  &  &  &  &  &  &  &  &  \\
  \hline
\end{tabular}
\label{table:FEA_Comparison}
\end{table*}


\begin{figure*}[b]
	\includegraphics[width=7in]{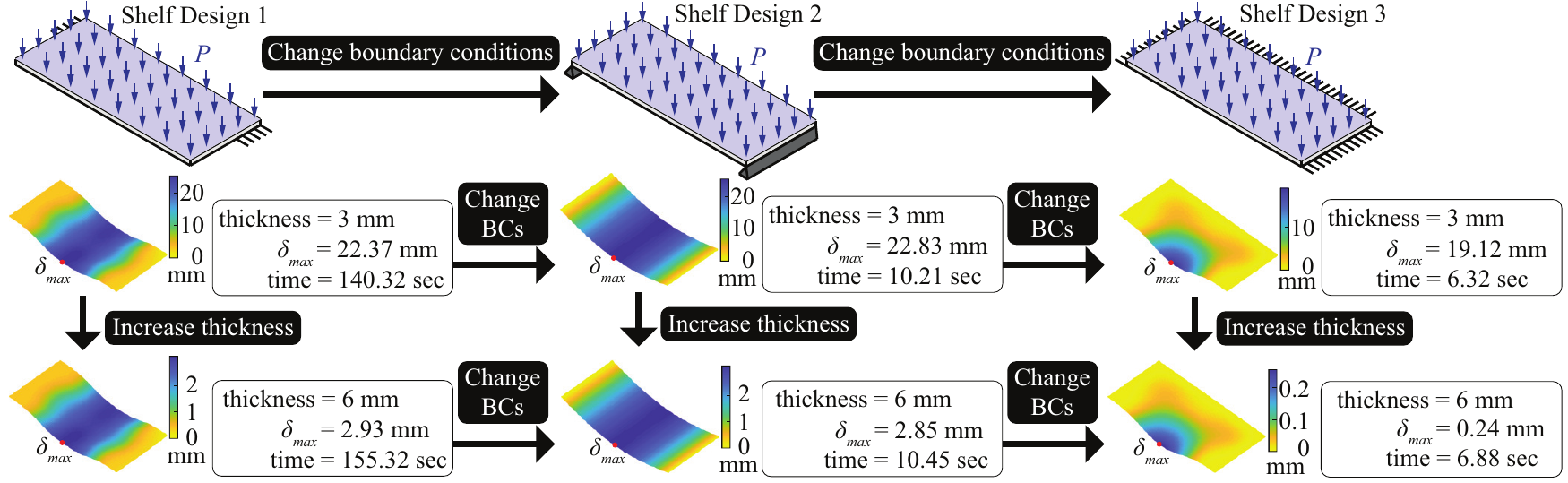}
	\centering
	\caption{Computational design of a bookcase. Boundary conditions and loading conditions are presented as schematic drawings. Analysis results (i.e. deflections) from our design tool are presented with corresponding thicknesses, maximum deflections and calculation time. Maximum deflection point marked in red dots.}
	\label{fig:Demo_Shelf}
\end{figure*}


\subsection{Computation Complexity}
\label{section:Computation_Complexity}

We use big O notation to denote the asymptotic complexity. We can observe that the dominating terms in the expression for the computational complexity of the algorithm come from the calculation of the coefficients of $c_i$ and $a_j$, calculated from solving Eq.~\ref{eq:omega} and \ref{eq: phi_p} respectively. Equation~\ref{eq: residual_calculation} produces a system of equations that allows us to solve Eq.~\ref{eq:residual}. Since solving a system of linear equations has the same order of complexity as matrix multiplication, we estimate that our algorithm has a computational complexity of 
\begin{equation}
    O(r^{\alpha} + k^{\alpha})
\end{equation}

With $k$ and $r$ consistent with their definitions in Eq.~\ref{eq:residual} and \ref{eq: phi_p} respectively. $\alpha$ is a constant that ranges from 2.3 to 3 depending on the method of matrix multiplication or matrix inversion used \cite{strassen_1969,6710599,coppersmith_winograd_1987}. For the purposes of this algorithm, \cite{6710599}, Gauss-Jordan elimination, or QR decomposition is generally most efficient due to the small constant term compared to \cite{strassen_1969} and \cite{coppersmith_winograd_1987} as $r$, $k$ should be relatively small ($r + k \leq 1000$). We can estimate for relatively small values of n ($n\approx 10$), that $k >> r$ from the data collected in Table~\ref{table:FEA_Comparison} (Section IV-B). In \cite{5744437}, FEA was found to have a computational complexity of 
\begin{equation}
    O(NW^2)
\end{equation}

\noindent where $W$ is the bandwidth of the stiffness matrix, and $N$ is the number of nodes, with the dominant term coming from the complexity of a linear solver. The bandwidth of the stiffness matrix, and the number of nodes are both inversely proportional to the mesh size. Since both algorithms rely on linear solvers for the dominant complexity term, the matrix needed for FEA to achieve the same accuracy compared to our algorithm is larger. This suggests that our algorithm also requires less storage than FEA.


\subsection{Calculation Speed and Accuracy}

To verify our analysis in Section~\ref{section:Computation_Complexity}, we compared calculation time complexity and accuracy. We compared results from FEA and those from our design tool compiled in MATLAB with one existing analytical solution \cite{imrak2007exact}. The computation machine we used was Intel\textsuperscript{\textregistered} Core\textsuperscript{TM} i7-7567U CPU @ 2.50GHz with 16.0 GB RAM.
The example plate problem used the following properties:

\begin{itemize}
    \item Geometry properties: \SI{250}{\milli\meter}$\times$\SI{500}{\milli\meter}$\times$\SI{0.1}{\milli\meter} rectangular plate;
    \item Mechanical properties: Young's Modulus $E = 2.1 \times 10^{2}\SI{}{G\pascal}$, Poisson's ratio $\nu = 0.3$;
    \item Boundary conditions: all four edges clamped;
    \item Loading condition: a $P=\SI{0.8}{\newton\per\meter^2}$ uniform load.
\end{itemize}


We compared the maximum deflections from FEA and our design tool with the analytical result \cite{zozulya2011numerical, qin1996transient} (Table~\ref{table:FEA_Comparison}). CPU time was also compared between FEA and our design tool.
Results in Table~\ref{table:FEA_Comparison} shows increasing polynomial order results in higher accuracy but longer computation time.
In FEA, we started with a large mesh size to achieve compatible solving time as that of our design tool and then decreased it to achieve compatible accuracy.
In the example case shown in Table~\ref{table:FEA_Comparison}, to achieve the same calculation speed, our design tool had higher accuracy (81.78\% error v.s. 1.35\% error and 71.08\% error v.s. 0.04\% error); to achieve competitive accuracy, our approach was at least 764\% faster. From these results, we have evidence for our hypothesis that our design tool requires a smaller matrix to be solved for the same accuracy compared to FEA. This demonstrates the efficiency of our design tool.






\section{Design Tool Demonstration}
\label{section:Design_Implementation}

When designing furniture, one important factor is to ensure the structural rigidity, which requires domain experts' knowledge and/or extensive modelings to correctly predict structural behaviors. Using the subsequent three design problems, we will prove the following capabilities and feasibilities of our design tool: (1) it is suitable for non-expert users to rapidly predict failure modes such as material breakage and structural over-deflection; (2) it is applicable to isotropic material (e.g. glass and plastic) and anisotropic material (e.g. wood); and (3) it is suitable for complex boundary conditions and loading conditions.



\subsection{Furniture}


\subsubsection{Bookcase}


Bookcases are usually assemblies of side panels with horizontal shelves, made out of orthotropic plywood \cite{eckelman1967furniture}.
To build a viable bookcase, one needs to determine whether shelves meet industrial qualification, which states $L/144$ ($L$ as the length of the shelf) to be the maximum acceptable deflection of a wood shelf \cite{standardunless}.

We decided to design and evaluate bookcase shelves to validate our design tool.
In our design, we chose to use Brich plywood as constructive material, which is a material can not be solved using existing intuitive solutions such as beam theory.
In our design, panels were assumed rigid in order to focus the design problems on shelves, which were $\SI{575}{\milli\meter} \times \SI{275}{\milli\meter}$ rectangular plates. This shelf geometry layout permitted a maximum $\SI{4}{\milli\meter}$ deflection.

In our analysis, we first evaluated a shelf design that had two shorter edges partially clamped (Shelf 1 in Fig.~\ref{fig:Demo_Shelf}). When applying $\SI{20}{\kilo\gram}$ weight across each $\SI{3}{\milli\meter}$ thick shelves, we noticed maximum deflection surpassed the maximum permitted deflection. We then adjusted boundary conditions and or design tool rapidly gave us an updated evaluation in approximately $\SI{10}{\second}$. We repeatedly and rapidly adjusted the boundary conditions for three designs and noticed none worked for $\SI{3}{\milli\meter}$ thick shelves. We then increased the thickness of each shelf and rapidly obtained the results which indicated all three shelf designs passed industrial qualification when shelf thickness is $\SI{6}{\milli\meter}$ (Fig.~\ref{fig:Demo_Shelf}).

\begin{figure}[t]
	\includegraphics[width=3.25in]{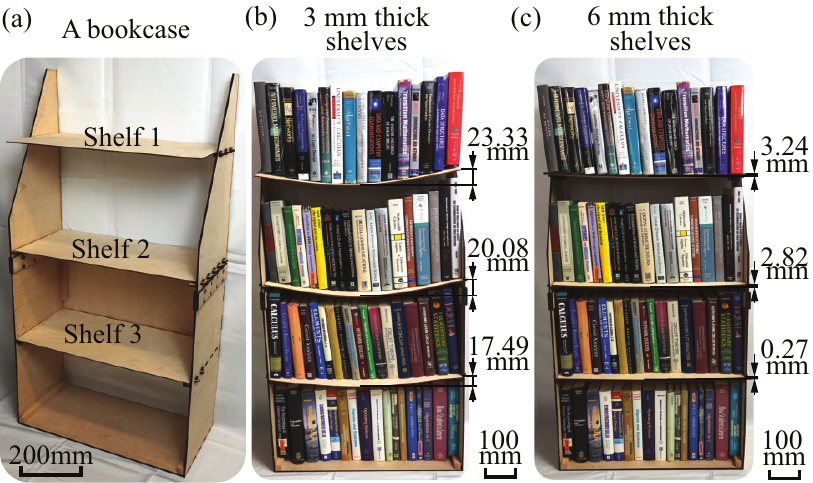}
	\centering
	\caption{A physical book case demonstration. (a) A physical bookcase with three shelves. (b) Loaded bookcase with $\SI{3}{\milli\meter}$ thick shelves. (c) Loaded bookcase with $\SI{6}{\milli\meter}$ thick shelves.}
	\label{fig:Demo_Shelf_Physical_Samples}
\end{figure}

To validate our design tool, we built two physical bookcases and applied $\SI{20}{\kilo\gram}$ weight across each shelf. Plates were cut using laser cutter (Trotec Speedy 300 Laser Engraver) and then assembled by nailing and slot-fitting \cite{saul2010sketchchair} depending on connection requirements (Fig.~\ref{fig:Demo_Shelf_Physical_Samples}(a)). $\SI{6}{\milli\meter}$ thick Birch plywood was used to build panels to ensure structural rigidity.
$\SI{3}{\milli\meter}$ thick shelves (Fig.~\ref{fig:Demo_Shelf_Physical_Samples}(b)) and $\SI{6}{\milli\meter}$ thick shelves (Fig.~\ref{fig:Demo_Shelf_Physical_Samples}(c)) were tested separately and results matched those in Fig.~\ref{fig:Demo_Shelf}(b) and (c).


\subsubsection{Glass Top Table}

\begin{figure}[!t]
	\includegraphics[width=3.25in]{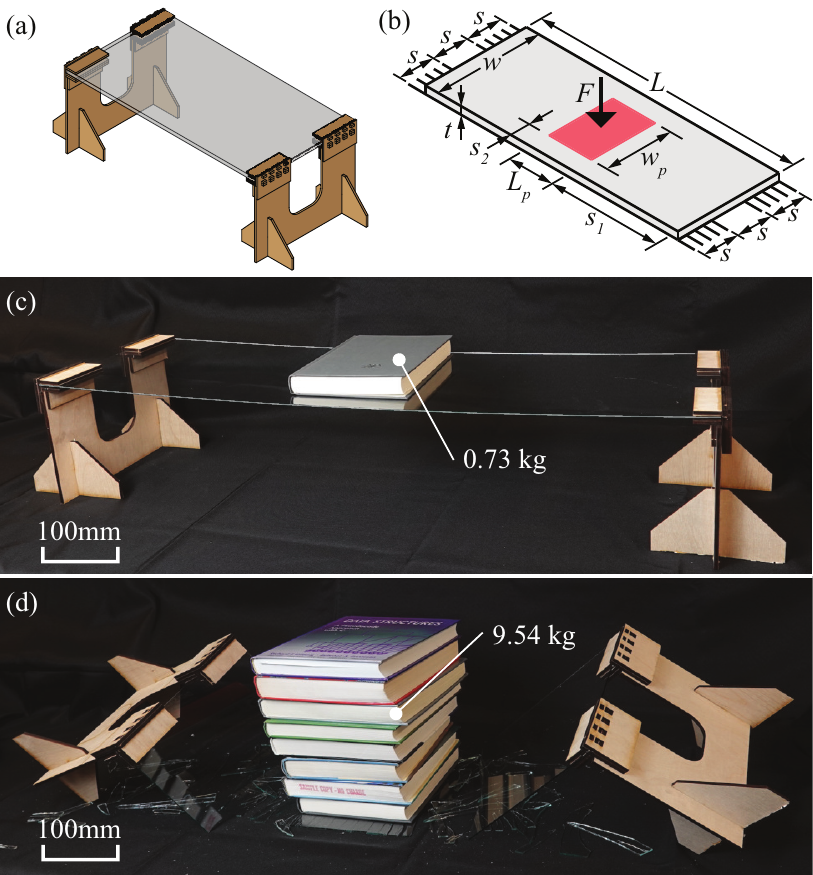}
	\centering
	\caption{Computational design of a glass top table. (a) Table design. (b) A schematic diagram for the glass top plate with boundary conditions and loading condition. (c) A $\SI{0.73}{kg}$ weight at the center with no damage on the table. (d) A $\SI{9.54}{kg}$ weight caused the glass to break. Parameters: $L = \SI{900}{\milli\meter}$, $w = \SI{300}{\milli\meter}$, $t = \SI{2.3}{\milli\meter}$, $L_p = \SI{160}{\milli\meter}$, $w_p = \SI{240}{\milli\meter}$, $s_1 = \SI{370}{\milli\meter}$, $s_2 = \SI{30}{\milli\meter}$ and $s = \SI{100}{\milli\meter}$.}
	\label{fig:Demo_Glass_Table}
\end{figure}

We also designed a rectangle glass top table with the four corners clamped (Fig.~\ref{fig:Demo_Glass_Table}(a)), as those commonly found in IKEA. 
The table was composed of a $\SI{900}{\milli\meter} \times \SI{300}{\milli\meter} \times \SI{2.3}{\milli\meter}$ clear glass top ($E=\SI{70}{GPa}$, $\nu=0.22$) and legs made out of flat-pack plywood assemblies. The glass top was partially clamped at the two shorter edges as shown in the schematic diagram in Fig.~\ref{fig:Demo_Glass_Table}(b).

We applied our design tool to evaluate the glass top. Based on the evaluation, when a $\SI{7.15}{kg}$ weight was displaced at the center as a $\SI{160}{\milli\meter} \times \SI{240}{\milli\meter}$ patch load, the maximum resulting stress on the plate would reach the glass's modulus of rupture ($\approx \SI{40}{MPa}$) and cause it to break. The entire evaluation took $\SI{23}{\second}$.


To validate our design tool, we built a glass table and first put a $\SI{0.73}{kg}$ book at the center (Fig.~\ref{fig:Demo_Glass_Table}(c)). There was no break detected on the glass, which matched the prediction from our design tool. We then increased the weight to and $\SI{6.77}{kg}$ without breaking the glass. When weight surpassed $\SI{7.15}{kg}$, glass top did break (Fig.~\ref{fig:Demo_Glass_Table}(d)) as expected from our design tool. This proved the accuracy of our design tool.

To demonstrate the necessity of our approach, we used beam theory to provide intuition.
When approximating the glass top as a beam fixed on the two ends with central distributed load \cite{gere1984mechanics}, results indicated the maximum stress to be $\SI{31.75}{MPa}$, which failed to predict the breakage.
This further indicates the necessity of our design tool non-expert users to both rapidly and correctly evaluate design with complex boundary conditions and loading conditions.

In sum, by presenting the above two furniture demonstration, our design tool has been proved to provide people without design experience an accessible technique to rapidly evaluate different designs and easily adjust parameters before spending time and effort building physical products.


\subsection{Origami Gripper}

\begin{figure}[t]
	\includegraphics[width=3.25in]{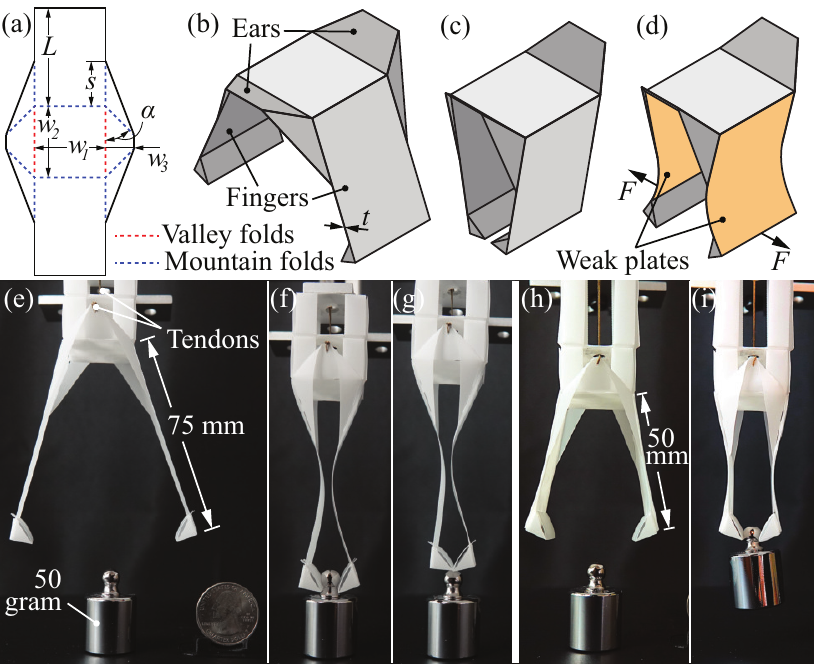}
	\centering
	\caption{Oriceps origami gripper. (a) Oriceps foldable pattern. (b) An open gripper. (c) A closed with No external force. (d) A deformed gripper under load. (e-g) Initial gripper failing to operate. (e) A gripper with $\SI{75}{\milli\meter}$ long fingers. (f) It tries to grasp a $\SI{50}{\gram}$ calibration weight. (g) It fails to lift the weight. (h-i) An improved gripper succeeding in operation. (h) A gripper with $\SI{50}{\milli\meter}$ long fingers. (i) It lifts the weight.}
	\label{fig:Gripper_Total}
\end{figure}

Oriceps origami gripper (Fig.~\ref{fig:Gripper_Total}(a)) is a potential solution for building disposable and inexpensive medical forceps as well as inexpensive toy robots \cite{edmondson2013oriceps, wilcox2015considering}. 
This gripper can open and close by pulling the ears to perform grasping operations (Fig.~\ref{fig:Gripper_Total}(b-c)).
However, due to the lack of bending rigidity on fingers, deformation on finger tips will reach operational limitation when gripper tries to lift heavy weights (Fig.~\ref{fig:Gripper_Total}(d)). Improvements on design need to be made on fingers to address this issue.


As shown in Fig.~\ref{fig:Gripper_Total}(e), a gripper (parameters in Table~\ref{table:gripper_iterative_design}) was built using Grafix Dura-Lar film as constructive material cut by Silhouette Cameo 4. It was actuated by pulling the ears using tendons. The gripper reached a $\SI{50}{\gram}$ calibration weight and closed to grasp the weight (Fig.~\ref{fig:Gripper_Total}(f)). As gripper tried to lift the weight, fingers bent and tips' deflection exceeded operation limits ($\SI{1}{\milli\meter}$) and failed to lift while other plates in the gripper still functioned (Fig.~\ref{fig:Gripper_Total}(g)).

\begin{figure}[b]
	\includegraphics[width=3.25in]{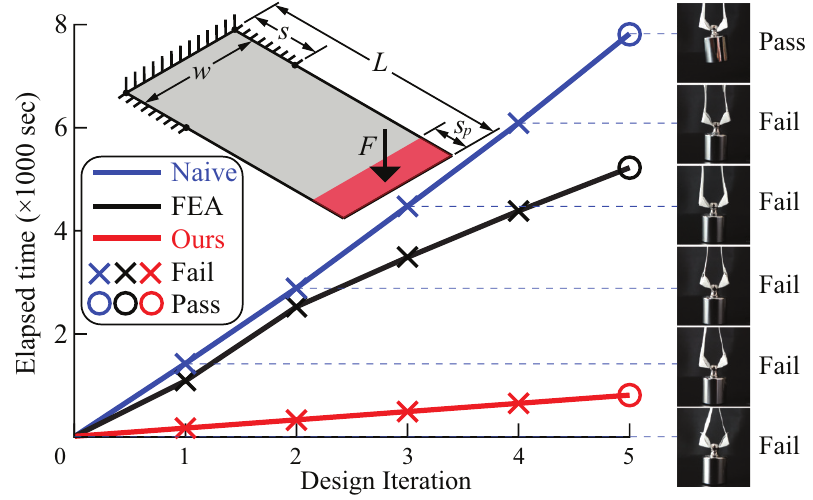}
	\centering
	\caption{Elapsed time comparison for the same design iteration but different approaches. $F = \SI{0.3}{\newton}$ and other parameters can be found in Table~\ref{table:gripper_iterative_design}.}
	\label{fig:Gripper_Design_Time}
\end{figure}

\begin{table}[t]
\centering
\caption{Oriceps origami gripper iterative design parameters and computation results. \\ ($w_1 = w_2 = s = \SI{25}{\milli\meter}$, $w_3 = \SI{15}{\milli\meter}$, $\alpha = \SI{50}{\degree}$.\\Italic text: failed designs. Bold text: successful design)}
\begin{tabular}{|c|c|c|c|c|c|}
\hline
\multirow{3}{*}{\begin{tabular}[c]{@{}c@{}}Design\\ iteration\end{tabular}} & \multirow{3}{*}{\begin{tabular}[c]{@{}c@{}}$t$\\ ($\SI{}{\milli\meter}$)\end{tabular}} & \multirow{3}{*}{\begin{tabular}[c]{@{}c@{}}$L$\\ ($\SI{}{\milli\meter}$)\end{tabular}} & \multicolumn{3}{c|}{Finger tip deflection} \\ \cline{4-6} 
 &  &  & \begin{tabular}[c]{@{}c@{}}Naive\\approach\end{tabular} & \begin{tabular}[c]{@{}c@{}}FEA\\ ($\SI{}{\milli\meter}$)\end{tabular} & \begin{tabular}[c]{@{}c@{}}Ours\\ ($\SI{}{\milli\meter}$)\end{tabular} \\ \hline
\textit{1} & \textit{0.37} & \textit{75} & \textit{Over-deflect} & \textit{28.20} & \textit{30.73} \\ \hline
\textit{2} & \textit{0.37} & \textit{62.5} & \textit{Over-deflect} & \textit{12.64} & \textit{16.64} \\ \hline
\textit{3} & \textit{0.55} & \textit{62.5} & \textit{Over-deflect} & \textit{3.89} & \textit{5.62} \\ \hline
\textit{4} & \textit{0.55} & \textit{50} & \textit{Over-deflect} & \textit{1.31} & \textit{1.45} \\ \hline
\textbf{5} & \textbf{0.76} & \textbf{50} & \textbf{Good-to-go} & \textbf{0.50} & \textbf{0.55} \\ \hline
\end{tabular}
\label{table:gripper_iterative_design}
\end{table}

In order to prove the efficiency of our design tool in an iterative design process, we compared it with other two common approaches: trial-and-error (naive) and FEA. It is worth noting that the analytical solution is not included here due to its limited applicability for simple geometry, boundary conditions, and loading.
We iteratively changed fingers' design parameters such as thickness $t$ and length $L$, fabricated and tested each design until a gripper was able to lift the weight (Fig.~\ref{fig:Gripper_Total}(h-i)).
Each design iteration took significant amount of time fabricating physical samples.
Example design parameters can be found in Table~\ref{table:gripper_iterative_design}, though other values and parameters can also be adjusted with the same manner.
Same finger designs, with boundary conditions and loading condition shown in Fig.~\ref{fig:Gripper_Design_Time}, were then evaluated using Solidworks Simulation and our design tool.
In Solidworks Simulation, every time finger geometry changed, meshing and calculation had to be re-computed which took significant amount of time and effort.
Same approach was applied to our design tool. In contrast, every time design parameters changed, our design tool would perform the evaluation directly after design input was updated.
Tip deflections are summarized in Table~\ref{table:gripper_iterative_design} and the comparison of the elapsed time for iterative design is shown in Fig.~\ref{fig:Gripper_Design_Time}.
It can be found that, on average, our approach is about 10 times more time-efficient than the naive approach and 8 times more than Solidworks Simulation during each iteration.

Moreover, to prove the necessity of our design tool, we applied back-of-the-envelope calculation commonly used by non-expert users as a comparison to prove intuitive approach would fail this task.
The finger was modeled as a cantilever beam with distributed line load on the tip using beam theory. Design 5 in Table~\ref{table:gripper_iterative_design} was applied to the calculation and a $\SI{1.6}{\milli\meter}$ tip deflection was obtained, which failed to capture the behavior and demonstrated the necessity of our approach.


\section{Conclusion and Discussion}
\label{section:Conclusion}

The presented design paradigm provides an insight in which 3D structures can be designed and evaluated.
Using our design tool, the planar surfaces in three dimensional structures can be rapidly and efficiently evaluated and redesigned without building physical samples. 
With our design tool, design iterations are now independent from long manufacturing time since every new design can be computationally evaluated within seconds.
The simplicity of our design tool gives non-expert users the ability to develop products with better quality and waste less material and time.



Future work include extending our design tool to efficiently analyze other plate behaviors such as out-of-plane twisting and in-plane buckling. A complete computational design system can be built based on this framework. We expect the complete system to be able to take any 3D structure made out of planar surfaces as input, decouple it into individual plates and apply analysis to individual plate so as to provide effective evaluation and further redesign recommendations.






\bibliographystyle{IEEEtran}
\bibliography{Reference}

\end{document}